\documentclass[sigconf]{acmart}

\usepackage{mathtools}
\usepackage{graphicx}
\usepackage{courier}
\usepackage{epstopdf}
\usepackage{color}
\usepackage{csquotes}

\usepackage{amsmath}
\usepackage{amsfonts}
\usepackage{amssymb}
\usepackage[subnum]{cases}
\usepackage{booktabs} 
\usepackage{pifont}

\usepackage{algorithm}
\usepackage{algorithmic}

\usepackage{amssymb}
\usepackage{pifont}

\usepackage {tikz}
\usetikzlibrary {positioning}
\definecolor {processblue}{cmyk}{0.96,0,0,0}
\usepackage{lipsum,adjustbox}

\usepackage[subnum]{cases}
\usepackage{color,soul}

\usepackage{subcaption}

\DeclareMathOperator*{\argmin}{argmin}

\usepackage{footnote}
\makesavenoteenv{tabular}

\usepackage{multirow}

\usepackage{amsmath, amssymb, bbm, xspace}

\def\spose#1{\hbox to 0pt{#1\hss}}

\def\text #1{\hbox{\quad#1\quad}}

\def\nthinsp{\mskip -2   mu}

\def\superstar{^{\raise 0.5pt\hbox{$\nthinsp *$}}}
\def\SUPERSTAR{^{\raise 0.5pt\hbox{$*$}}}

\def\lamstarT {\lambda^{\raise 0.5pt\hbox{$\nthinsp *$}T}}

\def\hbar{\skew{4.2}\bar h}

		\def\bk1{{\rm 1\kern-.17em l}}
		\def\bkD{{\rm I\kern-.17em D}}
		\def\bkR{{\rm I\kern-.17em R}}
		\def\bkP{{\rm I\kern-.17em P}}
		\def\bkY{{\bf \kern-.17em Y}}
		\def\bkZ{{\bf \kern-.17em Z}}

		\def\beq{\begin{eqnarray}}
		\def\bc{\begin{center}}
		\def\be{\begin{enumerate}}
		\def\bi{\begin{itemize}}
		\def\bs{\begin{small}}
		\def\bS{\begin{slide}}
		\def\ec{\end{center}}
		\def\ee{\end{enumerate}}
		\def\ei{\end{itemize}}
		\def\es{\end{small}}
		\def\eS{\end{slide}}
		\def\eeq{\end{eqnarray}}
		\def\qed{\quad \vrule height7.5pt width4.17pt depth0pt}

	\def\cp2problem#1#2#3#4{\fbox
		 {\begin{tabular*}{0.9\textwidth}
			{@{}l@{\extracolsep{\fill}}l@{\extracolsep{6pt}}l@{\extracolsep{\fill}}c@{}}
				#1 & & $#4 $
			\end{tabular*}}}

		\renewcommand{\emph}[1]{\textbf{#1}}

		\def\bk1{{\rm 1\kern-.17em l}}
		\def\bkD{{\rm I\kern-.17em D}}
		\def\bkR{{\rm I\kern-.17em R}}
		\def\bkP{{\rm I\kern-.17em P}}
		
		\def\bkZ{{\bf{Z}}}

\newcommand {\beeq}[1]{\begin{equation}\label{#1}}
\newcommand {\eeeq}{\end{equation}}
\newcommand {\bea}{\begin{eqnarray}}
\newcommand {\eea}{\end{eqnarray}}

\def\texitem#1{\par\smallskip\noindent\hangindent 25pt
               \hbox to 25pt {\hss #1 ~}\ignorespaces}

\usepackage{hyperref}
\usepackage{cleveref}[2012/02/15]

\crefformat{footnote}{#2\footnotemark[#1]#3}

\setcopyright{rightsretained}

\acmDOI{10.475/123_4}

\acmISBN{123-4567-24-567/08/06}

\acmConference[Conf'18]{}{XXXX}{XXXX} 
\acmYear{1997}
\copyrightyear{2016}

\acmArticle{4}
\acmPrice{15.00}

\begin{document}
\title[FSCNMF: Fusing Structure and Content for Information Network Representation]{FSCNMF: Fusing Structure and Content via Non-negative Matrix Factorization for Embedding Information Networks}

\author{Sambaran Bandyopadhyay}
\affiliation{\institution{IBM Research}}
\email{sambband@in.ibm.com}

\author{Harsh Kara}
\affiliation{\institution{Indian Institute of Science, Bangalore}}
\email{harshk@iisc.ac.in}

\author{Aswin Kannan}
\affiliation{\institution{IBM Research}}
\email{aswkanna@in.ibm.com}

\author{M N Murty}
\affiliation{\institution{Indian Institute of Science, Bangalore}}
\email{mnm@iisc.ac.in}





\begin{abstract}
Analysis and visualization of an information network can be facilitated better using an appropriate embedding of the network. Network embedding learns a compact low-dimensional vector representation for each node of the network, and uses this lower dimensional representation for different network analysis tasks. Only the structure of the network is considered by a majority of the current embedding algorithms. However, some content is associated with each node, in most of the practical applications, which can help to understand the underlying semantics of the network. It is not straightforward to integrate the content of each node in the current state-of-the-art network embedding methods. 

In this paper, we propose a nonnegative matrix factorization based optimization framework, namely FSCNMF which considers both the network structure and the content of the nodes while learning a lower dimensional representation of each node in the network. Our approach systematically regularizes structure based on content and vice versa to exploit the consistency between the structure and content to the best possible extent. We further extend the basic FSCNMF to an advanced method, namely FSCNMF++ to capture the higher order proximities in the network. We conduct experiments on real world information networks for different types of machine learning applications such as node clustering, visualization, and multi-class classification. The results show that our method can represent the network significantly better than the state-of-the-art algorithms and improve the performance across all the applications that we consider.
\end{abstract}

%
%



\settopmatter{printacmref=false} 

\maketitle

\section{Introduction}
Information networks are ubiquitous in our daily life and a variety of useful information can be extracted by mining them intelligently. Network analysis tasks such as node clustering (a.k.a, community detection), visualization, multi-class classification are some of the well-known problems dealt by the research community. All these tasks need a set of independent and informative features. Typically a network is represented in the form of a graph. Different types of raw representations like adjacency matrix or adjacency list have been used as a direct input to many machine learning algorithms. Unfortunately it is very difficult for machine learning algorithms to mine useful information from these representations as they belong to a very high dimensional space and also are highly sparse in nature. In the last few years, network embedding has emerged as a central topic of network analysis research. Network embedding maps the high dimensional networks to a low dimensional vector space such that the information loss is minimum in some sense, and also the features used in the representation are discriminative and complementary in nature. There are different types of network embedding methods that exist in the literature including Deepwalk \citep{perozzi2014deepwalk}, Line \cite{tang2015line}, node2vec \cite{grover2016node2vec}, TADW \cite{yang2015network}, and AANE \cite{huang2017accelerated}.

Learning network representation has various advantages over mining the network data directly. There is no need to vary the basic embedding learning algorithm to deal with different network analysis tasks. Most of the embedding methods are claimed to be generic so that existing network analysis methods for classification or clustering can perform well on these embeddings. Network embedding techniques are mostly unsupervised \cite{perozzi2014deepwalk} or semi-supervised \cite{huang2017label} in nature as it is common to encounter networks with a small number of labelled nodes or without any labelled node. Hence not much supervision is required to learn the embeddings. It is also possible to retain various network properties such as homophily, community structure \cite{wang2017community} in the resultant node embeddings.

These properties are naturally associated with each other. Homophily \cite{mcpherson2001birds} in networks characterizes communities based on similarity between the nodes involved. Even though such similarity is captured, typically,  based on structural properties of the network, combining the semantic content associated with the nodes can help in exploiting homophily better. From an optimization viewpoint better embeddings can be obtained by regularizing the objective based on structural data using content and vice versa. Besides, integrating content present in the nodes has been shown to be useful in various mining tasks such as unsupervised ranking of nodes in a graph \cite{hsu2017unsupervised}, user stance prediction \cite{dong2017weakly} and community detection \cite{liu2015community} in networks.

Unfortunately, most of the existing network embedding algorithms consider only the network structure such as nodes and the links while learning the low dimensional representation. But in most of the real-world networks, nodes are also associated with rich content in the form of text or images characterizing them. Naturally a network embedding is semantically incomplete without the use of such rich content. Lately there is some effort to fuse structure of the network with content to get better embeddings. We will briefly discuss them in the next section.

Our key contributions, in this paper, are:
\begin{itemize}
\item We propose a fast and unsupervised NMF based optimization framework, namely FSCNMF which considers both the network structure and the content of the nodes while learning a lower dimensional vector representation of each node in the network. In contrast to existing attributed network embedding literature, we use content as a regularizer over structure and vice-versa, which helps to exploit the coherence across the structure and the content of the nodes. Our algorithm outputs two regularized embeddings of the network corresponding to structure and content, which can be combined efficiently as the final representation.
\item We extend the basic FSCNMF to an advanced method, namely FSCNMF++ to capture the higher order proximities in the network. We also generalize our approach to the case when each node is associated with different types of content such as text, image, and video.
\item We conduct experiments on real world information networks for different types of network analysis tasks such as node clustering, network visualization and multi-class classification. The results demonstrate the superiority of our algorithms over the state-of-the-art approaches (up to \textbf{65.42\%} improvement in performance over the best of the baselines). The code of the algorithm has been made publicly available to ease the reproducibility of the results.
\end{itemize}

The rest of the paper is organized as follows. Section \ref{sec:rw} discusses the state-of-the art network embedding techniques and detects some research gap there. Next we formally present the problem description in Section \ref{sec:prob}. The details of the proposed optimization framework and the algorithms, and their analysis are done in Sections \ref{sec:sol}, \ref{sec:solvingOpt} and \ref{sec:scalability}. Section \ref{sec:FSCNMF++} proposes an extension of FSCNMF to capture the higher order proximities in the network and Section \ref{sec:multipleCon} generalizes the approach for different types of content.
Section \ref{sec:exp} experimentally shows the merit of our approach over the baseline algorithms on multiple real networks. Finally we conclude with the key observations in Section \ref{sec:con}.

\section{Related Work}\label{sec:rw}
In this section, we briefly discuss some of the important related work. Detailed survey on network embedding (NE) or Network Representation Learning (NRL) can be found in \cite{hamilton2017representation}.
Feature engineering techniques were there for networks for a long time. Conventional feature extraction methods use hand-crafted features based on the domain knowledge related to the networks \cite{gallagher2010leveraging}. Different linear and non-linear unsupervised dimensionality reduction approaches such as PCA \cite{wold1987principal} and ISOMAP \cite{bengio2004out} have also been used to map the network to a lower dimensional vector space.

\textbf{Sampling based Embedding}: Advancement of the representation learning for natural language modeling has motivated the researchers to adopt them in the domain of information networks. DeepWalk \cite{perozzi2014deepwalk} uses the idea of vector representation of words \cite{mikolov2013efficient} in a document to node representation in networks. A document can be represented by a sequence of words. Similar to that, a network can also be represented by a sequence of nodes. DeepWalk employs a uniform random walk from a node until the maximum length of the random walk is reached. Two different objective functions have been used to capture first and second order proximities respectively and an edge sampling strategy is proposed to solve the joint optimization for node embedding in Line \cite{tang2015line}. In node2vec \cite{grover2016node2vec}, authors have proposed a biased random walk to explore the diverse neighborhood while incorporating first and second order proximities in network representation. All these methods use different sampling strategies or different hyper parameters to obtain the best network embedding for a particular task. Hence it is difficult to find the best strategy which is consistent over different types of networks. 

\textbf{DL based Embedding}: There are also some network embedding techniques based on deep learning. In \cite{wang2016structural}, authors propose a structural deep network embedding method where they propose a semi-supervised deep model and second-order proximity is used by the unsupervised component to capture the global network structure. The idea of using convolutional neural networks for graph embedding has been proposed in \cite{niepert2016learning,kipf2016semi}. GCN with node attributes (GraphSage) has been proposed in \cite{hamilton2017inductive} with an inductive learning setup.  Deep learning architecture has also been used for heterogeneous networks \cite{chang2015heterogeneous}. 

\textbf{NMF based Embedding}: Nonnegative matrix factorization based network embedding techniques have also been explored in the literature to some extent. A network representation approach based on factorizing higher orders of adjacency matrix has been proposed in \cite{cao2015grarep}. In \cite{tu2016max}, authors have shown the equivalence of Deepwalk embedding technique to that of a matrix factorization objective, and further combined that objective to a max-margin classifier to propose a semi-supervised network embedding technique. Modularity maximization based community detection method has been integrated with the objective of nonnegative matrix factorization in \cite{wang2017community} to represent an information network. In \cite{yang2017fast}, authors have captured different node proximities in a network by respective powers of the adjacency matrix and proposed an algorithm to approximate the higher order proximities. 

\textbf{Incorporating Content in Embedding}: A major limitation in all of the above works is that they use only the network structure for embedding. But for most of the real-world networks, rich content information such as textual description is associated with the nodes. Fusing such content is not straightforward in any of the above approaches. There is some limited amount of work present in the literature to fuse structure with content for network representation. In \cite{yang2015network}, authors have presented a joint matrix factorization based approach (TADW) for fusing content and structure. But their framework directly learns one embedding from content and structure together. In case when there is noise or inconsistency between structure and content, such a direct approach is prone to be affected more. Extending the idea of TADW, a semi supervised approach for node embedding based on matrix factorization which incorporates empirical loss minimization on the labeled nodes has been proposed in \cite{zhang2016collective}. An attributed network embedding technique (AANE) is proposed in \cite{huang2017accelerated}. The authors have used symmetric matrix factorization to get embeddings from the similarity matrix over the attributes, and use link structure of the network to ensure the embedding of the two connected node is similar. Another semi-supervised attributed embedding is proposed in \cite{huang2017label} where the label information of some nodes are used along with structure and attributes. A social network embedding technique with incomplete user posts and contents is proposed in \cite{zhang2017user}. Compared to these existing works, we propose a novel unsupervised NMF based optimization framework, which uses content as a regularizer of structure and vice versa, and derives two representations, one for structure and the other for content, of a network by iteratively minimizing the distance between them.

\section{Problem Formulation} \label{sec:prob}
An information network is typically represented by a graph as $G = (V, E, F)$, where $V=\{v_1, v_2,\cdots, v_n\}$ is the set of nodes (a.k.a. vertexes), each representing a data object. $E \subset \{(v_i,v_j) | v_i,v_j \in V \}$ is the set of edges between the vertexes. 
Each edge $e \in E$ is an ordered pair $e = (v_i, v_j)$ and is associated with a weight $w_{v_i,v_j} > 0$, which indicates the strength of the relation. If $G$ is undirected, we have $(v_i, v_j) \equiv (v_j, v_i)$ and $w_{v_i,v_j} \equiv w_{v_j,v_i}$; if $G$ is unweighted, $w_{v_i,v_j} = 1$, $\forall (v_i,v_j) \in E$.
$F = \{f_i \;|\; i \in \{1,2,\cdots,n\} \}$, where $f_i \in \mathbb{R}^d$ is the word vector (content) associated with the node $v_i \in V$. For simplicity we assume the content to be only text in the rest of this paper, but a similar approach can be taken even if there are different types of content such as image or video.

Let us denote the $n \times n$ dimensional adjacency matrix of the graph $G$ by $A = (a_{i,j})$, where $a_{i,j}=w_{v_i,v_j}$ if $(v_i,v_j) \in E$, and $a_{i,j}=0$ otherwise. So $i$th row of $A$ contains the immediate neighborhood information for node $i$. Clearly for a large network, the matrix $A$ is highly sparse in nature.
Traditionally $F$ can be represented by a matrix $C$ based on bag-of-word models. In a bag-of-word model, 
typically stop words are removed, and stemming is done as a preprocessing step. Each row of this matrix is a tf-idf vector for the textual content at the corresponding node. So the dimension of the matrix $C$ is $n \times d$, where $d$ is the number of unique words (after the preprocessing) in the corpus.

Given $G$, the task is to find some low dimensional vectorial representation of $G$ which is consistent with both the structure of the network and the content of the nodes. More formally, for the given network $G$, the network embedding is to learn a function $f : v_i \mapsto \mathbf{y_i} \in \mathbb{R}^k$, i.e., it maps every vertex to a $k$ dimensional vector, where $k < min(n,d)$. The representations should preserve the underlying semantics of the network. Hence the nodes which are close to each other in terms of their topogrphical distance or similarity in content should have similar representation. 


\section{Solution Approach: FSCNMF}\label{sec:sol}

NMF has been shown to be effective to find the underlying semantics in a network. NMF based optimization has also been used in literature in the context of network embedding \cite{tu2016max,yang2015network}. In this paper, we propose an optimization framework based on NMF to fuse structure and content of a network with proper regularization. Here adjacency matrix $A$ is based on the structure of the network, whereas the content matrix $C$ is based on the textual content in each node of the graph. Hence in an ideal scenario, the representations found solely based on $A$ would match well with the representations found solely based on $C$. But in reality, they may differ due to noise and topological inconsistency in the network. But still there should be a strong semantic coherence between the two, and that is what we want to leverage in this framework.

\subsection{Learning from the Structure}
Given the $n \times n$ adjacency matrix $A$ based on structure of the network, we want to find a low rank approximation of the same. Hence $A$ can be factorized as $A \approx B_1 B_2$, where the dimension of the nonnegative matrices $B_1$ and $B_2$ are $n \times k$, $k \times n$ respectively, and $k << n$. This low rank matrix factorization has been shown to be useful as the direct representation (such as adjacency matrix) of a large network is extremely high dimensional and sparse in nature. Each row of the matrix $B_1$ can be regarded as the representation or embedding of the corresponding node. We minimize the Frobenius norm of the approximation error to get the matrices $B_1$ and $B_2$. Also to avoid overfitting, regularization terms with weight parameters $\alpha_2, \alpha_3 \geq 0$ are employed. Hence we have:
\begin{equation}\label{eq:struc}
  B_1, B_2 = \argmin_{B_1, B_2 \geq 0} || A - B_1 B_2 ||_F^2 + \alpha_2 ||B_1||_F^2 + \alpha_3 ||B_2||_F^2
\end{equation}

\subsection{Learning from the Content}
As mentioned before, content or attributes in individual nodes contains crucial information about the similarity of the node with the other nodes in the network. For example, reasonable accuracy has been obtained just by using the content of each paper to cluster the nodes in a citation network. So given the $n \times d$ content matrix $C$, we again approximate it by the lower rank matrices $U$ and $V$ by minimizing the Frobenius norm as follows: 
\begin{equation}\label{eq:content}
  U, V = \argmin_{U, V \geq 0} || C - UV ||_F^2  + \beta_2 ||U||_F^2 + \beta_3 ||V||_F^2
\end{equation}
Here $U$ and $V$ are two matrices of dimensions $n \times k$ and $k \times d$ respectively, with $\beta_2, \beta_3 \geq 0$ as the weight parameters. The rows of $U$ represent the embeddings of the corresponding node in the network, only based on the content.

\subsection{Fusing Structure and Content} \label{sec:FSCNMF}
Last two subsections described the individual learning from structure and content respectively. But in most of the information networks, link structure and content are highly correlated and they exhibit common properties like homophily in the network. Sometimes content drives the formation of links, and hence the link structure of the network. One intuitive way to generate a single embedding of the network by using both structure and content is to use joint non-negative matrix factorization by replacing $U$ with $B_1$ in Equation \ref{eq:content}. But that may not work in practice as large information networks are noisy and often there is significant inconsistency between structure and content. Hence using the same embedding matrix in both the cost functions is crude and can lead to very poor locally optimal solution. Instead of that, we propose FSCNMF which uses content as a regularizer over structure and vice-versa. 

So given an embedding matrix $U$ based on the content, we would like to obtain the embedding matrix $B_1$ based on the link structure by minimizing the following cost function:
\begin{equation}\label{eq:cost1}
    D_1(B_1,B_2) = || A - B_1 B_2 ||_F^2 + \alpha_1 ||B_1 - U||_F^2 + \alpha_2 ||B_1||_F^2 + \alpha_3 ||B_2||_F^2
\end{equation}
As we want to exploit the consistency between structure and content, the term $||B_1 - U||_F^2$ would try to pull $B_1$ close to $U$. The weight parameter $\alpha_1 \geq 0$ controls the importance of content while optimizing the embedding from structure.
Hence the updated matrices $B_1$ and $B_2$ can be obtained as follows. 
\begin{equation}\label{eq:B}
  B_1, B_2 = \argmin_{B_1, B_2 \geq 0} D_1(B_1, B_2) 
\end{equation}

Similarly, given an embedding matrix $B_1$ based on structure, embedding matrix $U$ based on content can be found by minimizing the cost function below:
\begin{equation}\label{eq:cost2}
    D_2(U,V) = || C - UV ||_F^2 + \beta_1 ||U - B_1||_F^2 + \beta_2 ||U||_F^2 + \beta_3 ||V||_F^2
\end{equation}
Again the term $||U - B_1||_F^2$ would not allow the content embedding matrix $U$ to deviate significantly from the given structure embedding matrix $B_1$. The weight parameter $\beta_1 \geq 0$ controls the importance of structure while optimizing the embedding from the content. 
Then the updated values of $U$ and $V$ are calculated as:
\begin{equation}\label{eq:UV}
  U, V = \argmin_{U,V \geq 0} D_2(U,V)
\end{equation}

We use the above two optimizations in Eq. \ref{eq:B} and \ref{eq:UV} multiple times in an iterative way to get the final embeddings of the network. The approach is briefed in Algorithm \ref{alg:FSCNMF}.
In the framework proposed above, one can easily incorporate the prior knowledge of the network quality and semantics. For example, if the content of the network is known to be more informative than the link structure, then one should give more importance to the initial representation in $U$ than that in $B_1$. This can be accomplished by setting a higher value for $\alpha_1$ than that of $\beta_1$. On the other hand, a higher value of $\beta_1$ gives more importance to the structure of the network than the content, and push the overall representation to be more consistent with the structure (see Section \ref{sec:sensitivity}). This flexibility between the structure and content was not present in most of the existing network embedding literature as discussed in Section \ref{sec:rw}.

At the end of the optimization, we get two different embeddings $B_1$ and $U$ of the network. There are some choices possible to take the final embedding.
\begin{itemize}
\item If the structure and content are consistent, the matrices $B_1$ and $U$ are likely to be similar. In that case any of the two matrices, or an convex combination of the two matrices in the form $\left( \gamma \times B_1 + (1-\gamma) \times U \right)$, $0 \leq \gamma \leq 1$ would be a good choice for the final representation of the network. We conduct detailed experiments in Section \ref{sec:exp}.
\item If there is a prior information available on the quality (informativeness and less noisy) of structure and content, one can choose the matrix accordingly. For example, if it is the case that only very few key words are available, for example, as the content for each paper in a citation network, whereas the link structure of the network is good, we can choose $B_1$ as the final representation of the network ($\gamma = 1$), rather than choosing $U$. In this case, the content embedding guides the evolution of structure embedding and vice versa, but we give more importance to structure than content. 
\end{itemize}

\begin{algorithm} 
  \caption{\textbf{FSCNMF} - Network Embedding Algorithm}
  \label{alg:FSCNMF}
\begin{algorithmic}[1]
      
	\STATE Input: The graph $G=(V,E,F)$, dimension of the embedding space $k < min(n,d)$
    \STATE Output: A combination $B_1$ and $U$ as the representation of $G$
	\STATE Calculate the adjacency matrix $A$ and the content matrix $C$ of the graph $G$
	\STATE Initialize the matrices $B_1, B_2, U, V$  
	\STATE \textbf{While (a termination condition is not satisfied)}
	\STATE Iterative approach to solve the optimization problem in Eq. \ref{eq:B} - Alternately update $B_1$ and $B_2$ as stated in Eq. \ref{eq:updateB_1} and \ref{eq:updateB_2} for a fixed number of times
	\STATE Iterative approach to solve the optimization problem in Eq. \ref{eq:UV} - Alternately update $U$ and $V$ as given in Eq. \ref{eq:updateU} and \ref{eq:updateV} for a fixed number of times
	\STATE \textbf{End While}
	\end{algorithmic}
  \end{algorithm} 

Next we discuss the solution for each of the above optimization problems in detail.

\section{Solving the Optimization Problems}\label{sec:solvingOpt}
We derive the necessary update rules to solve the optimization problems in Equation \ref{eq:cost1} and \ref{eq:cost2} in this section. 
It is to be noted that the cost function in Eq. \ref{eq:B} is convex with respect to $B_1$ when $B_2$ is assumed to be fixed, and vice versa. Similar observations can be made about the cost function in Eq. \ref{eq:UV}. Using alternating minimization techniques, we first equate the partial derivative of Eq. \ref{eq:cost1} w.r.t. $B_1$ to $0$, keeping $B_2$ fixed as follows.
\begin{align*}
\centering
& 2 (B_1 B_2 - A) B_2^T + 2 \alpha_1 (B_1-U) + 2 \alpha_2 B_1 = 0 \\
& B_1 (B_2 B_2^T + \alpha_1 I + \alpha_2 I) = AB_2^T + \alpha_1 U ; \;\; \text{$I$ is an identity matrix} \\
& \therefore B_1 = (AB_2^T + \alpha_1 U) (B_2 B_2^T + \alpha_1 I + \alpha_2 I)^{-1}.
\end{align*}
Noting that the matrix $B_2 B_2^T$ is positive semi-definite and $I$ is an identity matrix, it is obvious to note that the above inverse exists. To ensure the non-negativity of $B_1$, we further impose the following.
To ensure the element wise non-negativity of $B_1$, we set all the negative elements to $0$ to get the final update rule for $B1$ as:
\begin{equation}\label{eq:updateB_1}
B_1 = [(AB_2^T + \alpha_1 U) (B_2 B_2^T + \alpha I + \alpha_2 I)^{-1}]_+
\end{equation} 
Here for any matrix $X$, ${[X]_+}_{ij} = X_{ij}$ if $X_{ij} \geq 0$, and ${[X]_+}_{ij} = 0$, otherwise. Similarly we keep $B_1$ fixed and equating the partial derivative of Eq. \ref{eq:cost1} w.r.t. $B_2$ to $0$ as shown below.
\begin{align*}
\centering
& 2 B_1^T (B_1 B_2 - A) + 2 \alpha_3 B_2 = 0 
\end{align*}
\begin{equation}\label{eq:updateB_2}
\therefore B_2 = [(B_1^t B_1 + \alpha_3 I)^{-1} B_1^T A]_+
\end{equation}
Similarly we get the following update rules for $U$ and $V$ as shown below.
\begin{equation}\label{eq:updateU}
U = [(CV^T + \beta_1 B_1) (V V^T + \beta_1 I + \beta_2 I)^{-1}]_+
\end{equation}
\begin{equation}\label{eq:updateV}
V = [(U^t U + \beta_3 I)^{-1} U^T C]_+
\end{equation}
The above update rules are used in Algorithm \ref{alg:FSCNMF}. 

\section{Scalability of FSCNMF}\label{sec:scalability}
In this section, we analyze the time complexity and scalability of FSCNMF. We assume that the matrix $A$ or $C$ are highly sparse in nature. Let us consider Eq. \ref{eq:updateB_1} first. Computation of $(AB_2^T + \alpha_1 U)$ would take a time of $O(nk)$ as A is sparse. The dimension of the matrix $(B_2B_2^T + \alpha_1 I + \alpha_2 I)$ is $k \times k$, and typically for most of our experiments, we keep the value of $k$ lesser than 100. Hence the total time to compute the update rule for $B_1$ is $O(nk + nk^2 + k^3) = O(nk^2)$. Similarly the runtime complexity to update $B_2$, $U$ and $V$ are $O(nk^2)$, $O(nk + dk^2)$ and $O(nk^2 + dk)$. So the runtime of FSCNMF is linear in the number of nodes of the network.

Moreover, all the update rules of the working variables in the algorithm are given in closed matrix forms. As the rows of any of those matrices are getting updated independently, one can use a distributed set up (similar to \cite{huang2017accelerated}) to update the matrices where different rows can be updated in parallel. This way we can make the algorithms even more faster without any loss of performance.

\section{FSCNMF++}\label{sec:FSCNMF++}
There are different types of proximities that exist in a network as stated in \cite{tang2015line}. For example, first order proximity captures the local pairwise proximity between the two nodes. It is characterized by the weight of the edge connecting them. Similarly, second order proximity between two nodes captures the similarity between their respective immediate neighborhoods. Extending that, any higher order proximity between two nodes in a network can be defined by the similarity of their respective higher order neighborhoods.
Higher order proximity information can also be useful for learning the network embeddings as it somehow captures the global behavior of the connectivity in a network.

FSCNMF approach discussed above only uses adjacency matrix to leverage the structural information present in the network. It is easy to fetch the $1$st order neighborhood (or proximity) information of any node from the adjacency matrix. But it might be difficult for a machine learning algorithm to fetch the higher order proximity information from the adjacency matrix directly. So instead of working with the adjacency matrix, we propose to use different powers of the adjacency matrix as follows.

It is known that different powers of adjacency matrix give global connectivity information about the network. For example, $(i,j)$th entry of $A^l$ gives the number of paths\footnote{It is to be noted that a node can appear multiple times within such a path.} of length $l$ between the nodes $i$ and $j$. The use of higher order powers of the adjacency matrix can also be motivated from the equivalence of Deepwalk and matrix factorization as shown in \cite{yang2015network,tu2016max}. Because of these two reasons, we propose FSCNMF++, which uses the following matrix instead of adjacency matrix in Algorithm \ref{alg:FSCNMF}.
\begin{align}\label{eq:M}
M = \frac{A + A^2 + \cdots + A^m}{m}
\end{align}
We will call the above version as the FSCNMF++ of order $m$, where $m \in \{1,2,\cdots\}$. As most of the real-world networks are highly sparse, i.e., $\mathcal{O}(E)=\mathcal{O}(V)$, computing the matrix $M$ requires $\mathcal{O}(n^2)$ time. Similarly for sparse networks, complexity of matrix factorization with squared loss is proportional to the number of non-zero elements in the matrix $M$ \cite{yu2014large}. Experimental results, provided in Section \ref{sec:exp}, demonstrate the superiority of FSCNMF++ for most of the machine learning applications on the networks.



\section{Generalization to Multiple Types of Content}\label{sec:multipleCon}
It is getting common in different social networks to have multiple types of content available in each node. For example, pages in Wikipedia have textual description and multiple images to describe the full content. Similarly, a Facebook profile has textual, image and video content, all of which are important to understand the semantic behavior of the user or the entity of interest. As these types of information are inherently different, it is not appropriate to concatenate everything as a single attribute vector for each node. Rather they need to be counted separately and combined at a later stage to get the embedding of the network. We can easily extend the proposed FSCNMF for such multiple types of content as follows.

As before, we want to approximate $M$ (or $A$) $\approx B_1 B_2$. Assume there are $T$ number of different types of content available with each node in the network. A corresponding content matrix is denoted by $C_t$, with dimensions $n \times d_t$, $\for t=1,2,\cdots,T$. Again we want to approximate $C_t$ by lower rank matrices $U_t$ and $V_t$ of dimensions $n \times k$ and $k \times d_t$ respectively such that $C_t \approx U_t V_t$. To get embedding of the network, we propose the following set of optimizations:
\begin{equation*}
    \argmin_{B_1, B_2 \geq 0} || M - B_1 B_2 ||_F^2 + \alpha_1 ||B_1||_F^2 + \alpha_2 ||B_2||_F^2 + \sum\limits_{t=1}^T \alpha_{3t} ||B_1 - U_t||_F^2
\end{equation*}
And for each content type $t$, $\forall t = 1,\cdots, T$:
\begin{align*}
 \argmin_{U_t, V_t \geq 0} \; & || C_t - U_t V_t ||_F^2  + \beta_1 ||U_t||_F + \beta_2 ||V_t||_F \\ 
  &+ \beta_3 || U_t - B_1 ||_F^2 + \sum\limits_{t' \neq t} \beta_{3t'} || U_t - U_{t'}||_F^2
\end{align*}
Here embedding from the link structure is regularized by the different types of content, and embedding from each type of content is regularized by structure and the other types of content in the network. Final embedding of the network can be derived by taking an average of the individual embeddings or by a biased approach as discussed in Section \ref{sec:sol}. Updates rules similar to Section \ref{sec:solvingOpt} can also be derived for all the variables in this case.

\section{Experimental Evaluation}\label{sec:exp}
In this section, we first discuss the datasets and the baseline algorithms used for the experiments. We conduct different types of experiments on the embeddings and analyze the results in detail.

\begin{table*}
    \caption{Summary of the datasets used: All the datasets used have both network structure and textual content in each node. \textit{\#Distinct Words} counts the number of distinct words in a network after eliminating the stop words and rare words from the whole content. \textit{Node Distribution} is the proportion of different communities in a dataset. \textit{Inter/intra links} is the ratio between the number of inter community links to that of the intra community links in a dataset.}
	\centering
	\begin{tabular}{*7c}
	\toprule
	\sffamily{Dataset} & \#Nodes & \#Edges & \#Labels & \#Distict Words & Node Distribution & Inter/Intra links \\
    \hline
	\midrule
    \sffamily{Citeseer} & 3312 & 4715 & 6 & 3703 & 0.18:0.08:0.21:0.20:0.18:0.15 & 0.34 \\
    \sffamily{Pubmed-Diabetes} & 19717 & 44338 & 3 & 500& 0.21:0.40:0.39 & 0.25\\
    \sffamily{Microsoft Academic Graph} & 30101 & 205654 & 3 & 6785 &0.51:0.21:0.28 & 0.06\\
    \sffamily{Wikipedia} & 40101  & 1806377  &  4 & 8722 & 0.24:0.31:0.27:0.18 & 0.23\\
\bottomrule
	\end{tabular}
	\label{tab:data}
	\end{table*} 

\subsection{Data Sets}
Following are the datasets used in this paper:

\textbf{Citeseer}\footnote{\label{footnote:dataurl}\url{https://linqs.soe.ucsc.edu/data}}: It consists of publications from Citeseer digital library. The nodes (publications) are divided into 6 categories, with the citations among them as the edges.

\textbf{Pubmed-Diabetes}\cref{footnote:dataurl}: It has scientific publications from the PubMed database consisting of 3 classes  of Diabetes. The edges denote the citations among them.

\textbf{Microsoft Academia (MSA)}: The MSA dataset contains scientific publication records and the citation relationship between those publications \cite{sinha2015overview}. Each publication has its title and some key phrases present in the dataset, which we have considered as the contents associated with the nodes. We have selected papers which appear in one of the following categories in computer science - \textit{Data Science}, \textit{Systems} and \textit{Theory}. We also filter out papers which have less than seven links (sum of incoming and outgoing links) in the original network.

\textbf{Wikipedia}: It is a directed network consisting of a collection of wikipedia articles from the 4 categories namely \textit{Sports}, \textit{Politics}, \textit{Music} and \textit{Medicine} with the edges representing the links between the articles.
Since a page can belong to multiple categories, we labeled the page's category as the main category which is at the least distance from it.
For each article, only the \textit{Introduction} text was taken as the content in the dataset.

A detailed specification of the datasets is given in Table \ref{tab:data}. All the datasets that we use have only text as the content. So we leave the experiments on multiple types of content for future work.

\subsection{Baseline Algorithms Used}
We compare the performance of our algorithms against the following diverse set of state-of-the-art approaches: DeepWalk \cite{perozzi2014deepwalk}, LINE \cite{tang2015line} (advanced version by considering
both the 1st and 2nd order of node proximity), node2vec \cite{grover2016node2vec}, TADW \cite{yang2015network}, AANE \cite{huang2017accelerated} and GraphSAGE \cite{hamilton2017inductive}. Among these baselines, DeepWalk, LINE and node2vec use only network structure, while TADW, AANE and GraphSAGE use both structure and the content of the nodes for generating the embeddings. We use the default settings (as applicable) given in the publicly available implementations of the respective baselines.

\subsection{Experimental Setup and Network Embedding}
In this section, we examine the behavior of the optimization algorithms of FSCNMF on real-world datasets. There are few parameters of FSCNMF to be set before the experiments. Following are the default parameter values that we set for all the experiments if not mentioned otherwise. Assuming there is no prior information available on the quality of structure and content, we set each of $\alpha_1$, $\alpha_2$, $\alpha_3$, $\beta_1$, $\beta_2$ and $\beta_3$ to be $1$. We observed that these values give similar weight to each component of the respective criterion function in optimizations (Eq. \ref{eq:cost1} and \ref{eq:cost2}). As a thumb rule, we set the dimension of the embedding subspace as $k$ $=$ $10 * (\#communities)$ for a dataset, assuming the number of communities to be known beforehand. We varied the values of some of these parameters in Section \ref{sec:sensitivity} to get further insight of the proposed algorithms. We use NMF with nonnegative double singular value decomposition as the initialization method for the matrices $B_1$, $B_2$, $U$ and $V$ (Section \ref{sec:sol}).

\begin{figure*}[h]
\centering

\begin{subfigure}{0.3\textwidth}
\includegraphics[scale=0.22]{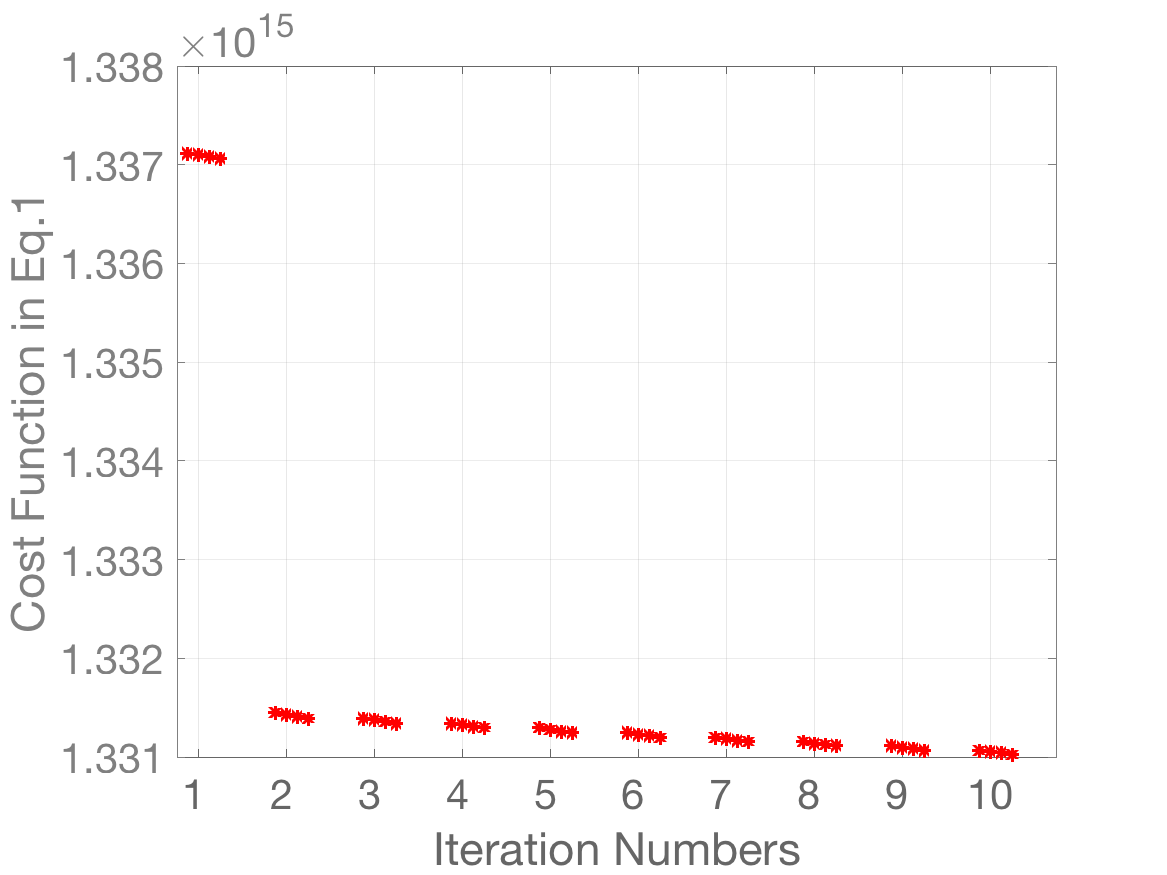}
\caption{Optimization in Eq \ref{eq:cost1} for Wikipedia}
\label{fig:opti1}
\end{subfigure}
\begin{subfigure}{0.3\textwidth}
\includegraphics[scale=0.22]{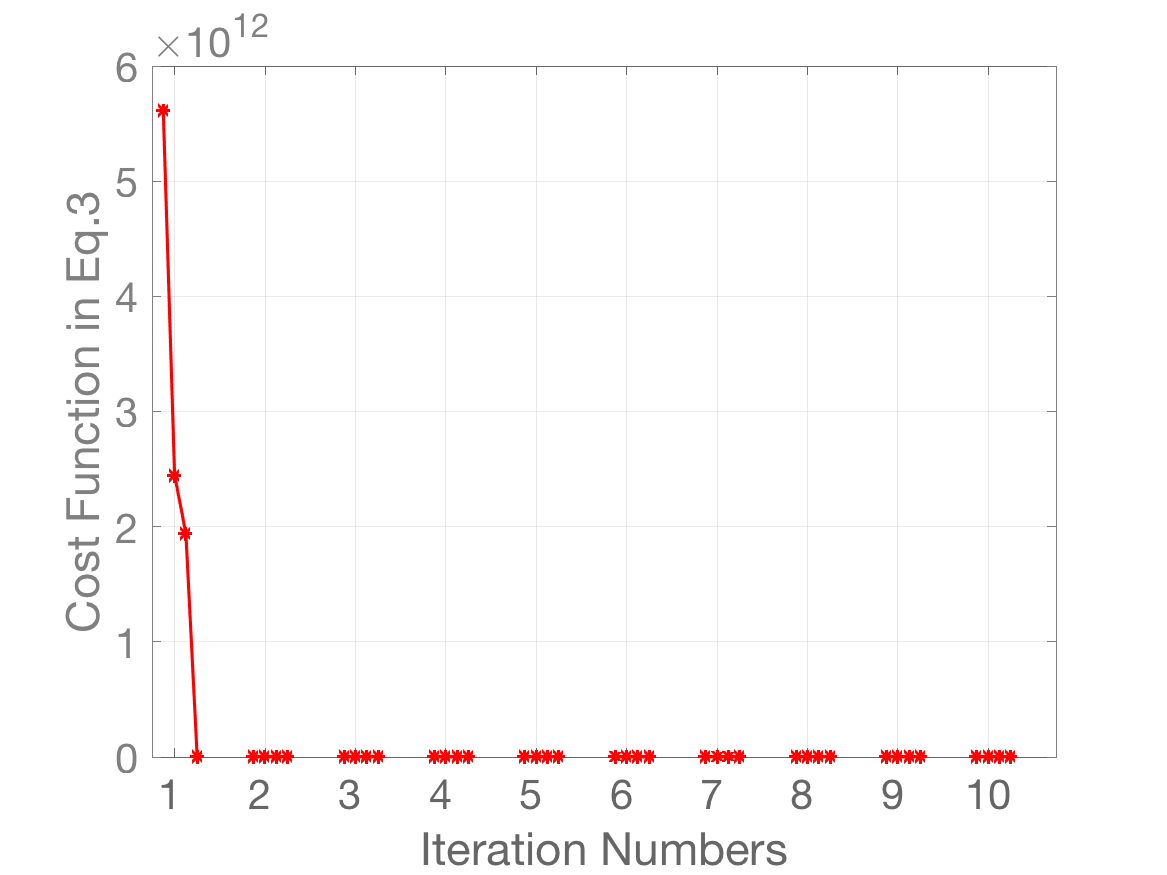}
\caption{Optimization in Eq \ref{eq:cost2} for Wikipedia}
\label{fig:opti2}
\end{subfigure}
\begin{subfigure}{0.3\textwidth}
\includegraphics[scale=0.22]{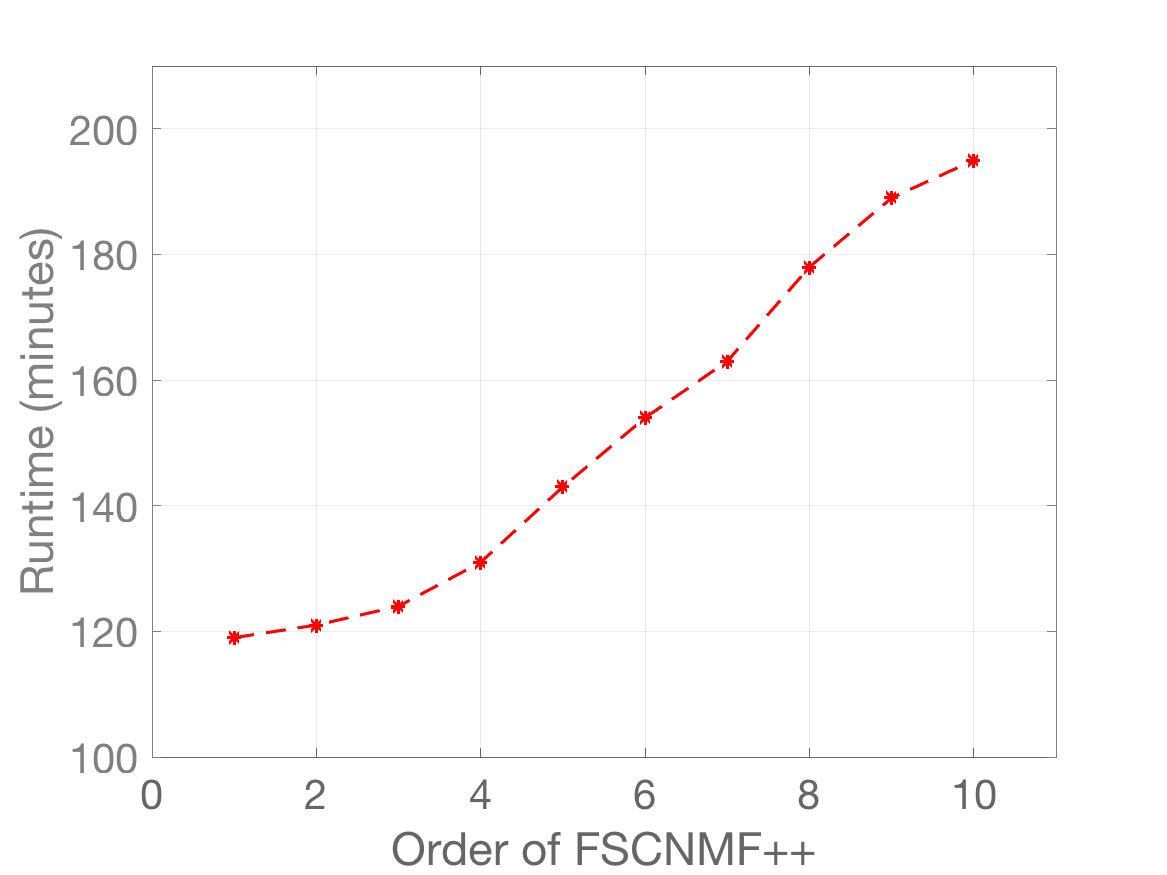}
\caption{Runtime for FSCNMF++ on Wikipedia}
\label{fig:runtime}
\end{subfigure}

\caption{Performance of FSCNMF++ of order 7 on the Wikipedia dataset: Plot (a) shows the updation of $B_1$ and $B_2$ with respect to the cost funtion in Equation \ref{eq:cost1}. Similarly, plot (b) depicts the updation of $U$ and $V$ with respect to the cost function in Equation \ref{eq:cost2}. The inter-dependencies between $B_1$, $B_2$ and $U$, $V$, as explained in Algo. \ref{alg:FSCNMF} (Steps 5-8) can be understood using plots (a) and (b). Between any two consecutive segments in Plot (a), the drop in the respective cost function is because of the update of $U$ and $V$ as depicted in Plot (b), and vice-versa. (c) The runtime of different orders of FSCNMF++ on the Wikipedia dataset. The plots depict the advantage of using fast sparse matrix multiplication to compute the matrix $M$ in Eq. \ref{eq:M}} 
\label{fig:optiandrun}
\end{figure*}

First we use Algorithm \ref{alg:FSCNMF} for FSCNMF and FSCNMF++. As described in Section \ref{sec:solvingOpt}, we proposed alternating minimization based update rules to optimize each cost function. We show the convergence of the proposed technique in Figures \ref{fig:opti1} and \ref{fig:opti2}. Both the plots depict the fast convergence of FSCNMF++ on Wikipedia dataset. We can also see the significant drop of the cost after outer iteration 1 in Figure \ref{fig:opti1}. This happens because the updated content matrix $U$ from the optimization in Eq. \ref{eq:UV} makes the cost in Eq. \ref{eq:cost1} go down. Thus we are able to successfully exploit the coherence of structure and content through the proposed optimization framework.

For FSCNMF of order $l$, the adjacency matrix $A$ needs to be raised up to the power of $l$. We use sparse matrix multiplication available in SciPy \cite{jones2014scipy} to compute different powers of $A$. The runtime for different orders of FSCNMF++ to generate the embeddings on the Wikipedia dataset is shown in Figure \ref{fig:runtime}. We observe that the difference between the runtime of any two consecutive orders of FSCNMF++ is in the range of few minutes.


\begin{table*}
\caption{Accuracy (\%) of node clustering by Spectral Clustering for different embedding methods}
	\centering
	\begin{tabular}{*9c}
	\toprule
	\sffamily{Dataset} & DeepWalk & LINE & node2vec & TADW & AANE & GraphSAGE & FSCNMF & FSCNMF++  \\
    \hline
	\midrule
    \sffamily{Citeseer} & 23.46 & 23.12 & 22.40 & 25.03 & 39.25 &  27.98 & 43.68 & \textbf{49.42} (Order=2) \\
    \sffamily{Pubmed-Diabetes} & 41.94 & 40.87 & 43.50 & 44.66 & 44.66 & 50.71   & 56.86 & \textbf{63.68} (Order=9) \\
    \sffamily{MSA} & 43.75  & 50.49  & 49.16  & 39.75 & 49.71 & 35.06  & 61.34 & \textbf{83.52} (Order=9)\\
    \sffamily{Wikipedia} & 31.90  &  30.57 & 37.03 & 38.90 & 45.78 & 66.06   & 45.12 & \textbf{71.66} (Order=7)\\
\bottomrule
	\end{tabular}
	\label{tab:cluster}
	\end{table*} 

\subsection{Applications to Node Clustering}\label{sec:clustering}
Node clustering is an important unsupervised machine learning task for network analysis. Here we check the performance of the proposed embedding technique for unsupervised discrimination of the individual classes in the dataset. To perform node clustering, we first get the vector embedding of each node using an embedding method. Then we apply spectral clustering \cite{ng2002spectral} on the vector representations to get the clusters in each case. We use this for all the embedding algorithms and on each of the datasets specified in Table \ref{tab:data}. As each node has a ground truth label in the datasets, we use unsupervised accuracy \cite{xie2016unsupervised} to judge the performance of clustering. Following is the definition of the unsupervised clustering accuracy:
$Acc(\mathcal{\hat{C}},\mathcal{C}) = \max_{\mathcal{P}} \frac{\sum\limits_{i=1}^n \mathbf{1}(\mathcal{P}(\mathcal{\hat{C}}_i), \mathcal{C}_i ) }{n}$.
Here $\mathcal{C}$ is the ground truth labeling of the dataset such that $\mathcal{C}_i$ gives the ground truth label of $i$th data point. Similarly $\mathcal{\hat{C}}$ is the clustering assignments discovered by some algorithm, and $\mathcal{P}$ is a permutation on the set of labels. As clustering is unsupervised and hence we do not know the exact mapping between the set of labels in the ground truth and that in the clustering, we have to consider all the permutations of the labels and select the accuracy corresponding to the best permutation. We assume $\mathbf{1}$ to be the identity function on $\mathbb{R}^2$, defined as $\mathbf{1}(a,b)=1$ if $a=b$ and $\mathbf{1}(a,b)=0$ if $a \neq b$.

We have compared the performance of the proposed embedding methods FSCNMF and FSCNMF++ for node clustering against all the baseline methods. For each embedding, the configurations in the spectral clustering is kept the same. The results are shown in Table \ref{tab:cluster}. FSCNMF++ with different orders turns out to be the best among all the embedding methods. FSCNMF remains to be the second best, except for the Wikipedia dataset where GraphSAGE outperforms FSCNMF. For MSA, only structure based embedding methods such as DeepWalk, Line and node2vec perform better than TADW and FSCNMF. This implies that content of the nodes does not play a significant role for the node clustering in MSA. This can be explained by the statistics mentioned in Table \ref{tab:data}. In MSA the ratio between the number of links between the communities to the number of links within the communities is only 0.06. This means the communities are well defined by the link structure in the network. Also the content within a node in MSA dataset contains only the title of the paper and few keywords. Hence TADW and AANE suffer as they use content in a rigid way, and the content was noisy in this case. Interestingly, due to the inherent flexibility of FSCNMF optimization formulation, both FSCNMF and FSCNMF++ are able to outperform all the baselines significantly even for the MSA dataset, when the content is noisy. For example, the clustering accuracy of FSCNMF++ is \textbf{65.42\%} better than that of LINE (best among all the baselines for clustering MSA dataset). 

\begin{figure}[h]
\centering
\includegraphics[scale=0.28]{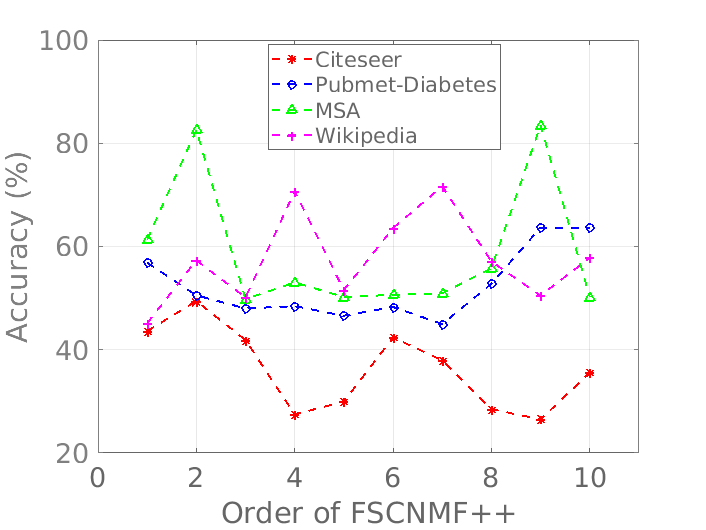}
\caption{Accuracy of clustering for different orders of FSCNMF++ on all the datasets.}
\label{fig:clustAcc}
\end{figure}

An important observation is that the spectral clustering algorithm on the embedding given by FSCNMF is faster by a factor of more than 100 compared to that on DeepWalk and LINE embeddings. We further investigate the effect of different orders of FSCNMF++ on node clustering in Figure \ref{fig:clustAcc}. It can be observed that for a smaller dataset like Citeseer, higher orders of FSCNMF++ do not improve the clustering accuracy much. In other words, higher order proximities can actually add noise to the resultant embeddings of a smaller network. Whereas for most of the large networks, clustering accuracies more or less improve up to a certain order of FSCNMF++ and then start decreasing\footnote{We observe the same phenomenon beyond order 10 for Pubmed-Diabetes dataset.}.

\begin{figure*}[h]
\centering

\begin{subfigure}{0.18\textwidth}
\includegraphics[scale=0.25]{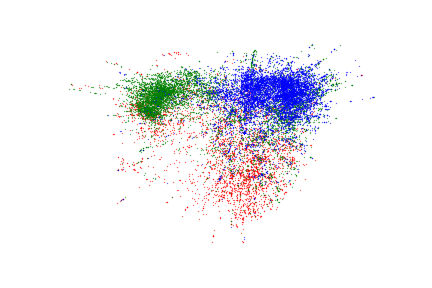}
\caption{DeepWalk}
\label{fig:subim1}
\end{subfigure}
\begin{subfigure}{0.18\textwidth}
\includegraphics[scale=0.25]{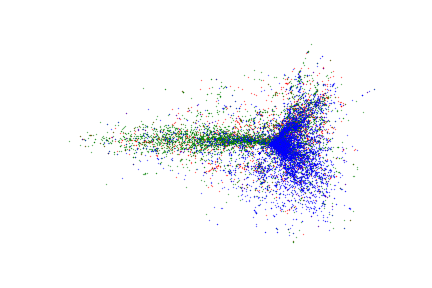}
\caption{Line}
\label{fig:subim1}
\end{subfigure}
\begin{subfigure}{0.18\textwidth}
\includegraphics[scale=0.25]{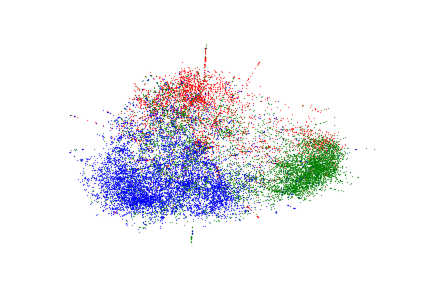}
\caption{node2vec}
\label{fig:subim1}
\end{subfigure} 
\\
\begin{subfigure}{0.18\textwidth}
\includegraphics[scale=0.25]{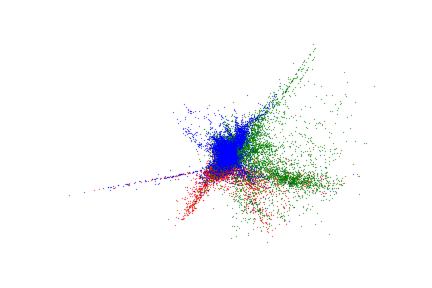}
\caption{TADW}
\label{fig:subim1}
\end{subfigure}
\begin{subfigure}{0.18\textwidth}
\includegraphics[scale=0.25]{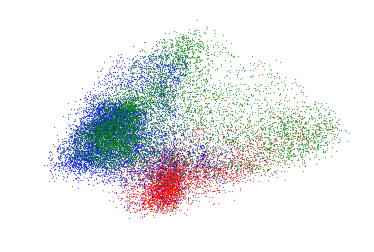}
\caption{AANE}
\label{fig:subim1}
\end{subfigure}
\begin{subfigure}{0.18\textwidth}
\includegraphics[scale=0.25]{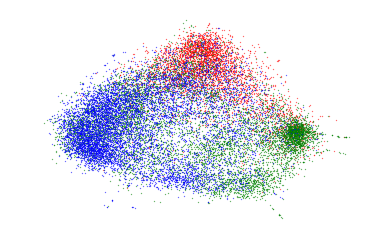}
\caption{Graphsage}
\label{fig:subim1}
\end{subfigure}
\begin{subfigure}{0.18\textwidth}
\includegraphics[scale=0.25]{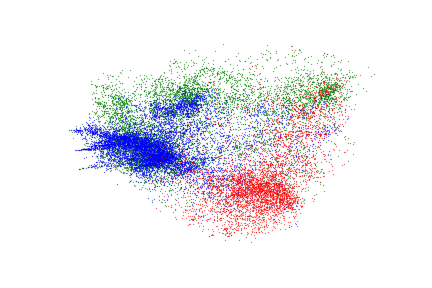}
\caption{FSCNMF++}
\end{subfigure}

\caption{Visualization of Pubmed-Diabetes dataset. Each point in the plot represents a node in the dataset. Each color corresponds to a particular group in the dataset.}
\label{fig:visuaPub}
\end{figure*}

    
\subsection{Applications to Network Visualization}
In network visualization or graph visualization, the goal is to map the network or its representations to a $2D$ space, and show if the plot in the $2D$ space discriminates different classes or communities present in the network. Network visualization is unsupervised as the labels of the nodes are not being used for learning the map, they can be used in the 2D plots for the better understanding of the quality of visualization. We use ISOMAP \cite{tenenbaum2000global} toolkit present in python Scikit-learn \cite{pedregosa2011scikit} library to convert the network embedding to 2D space.
We use same color for the nodes which belong to the same community, and different colors for the different communities. So a good visualization is that where nodes in the same community are near each other and nodes from different communities are separated from each other.

\begin{figure*}[h]
\centering

\begin{subfigure}{0.18\textwidth}
\includegraphics[scale=0.25]{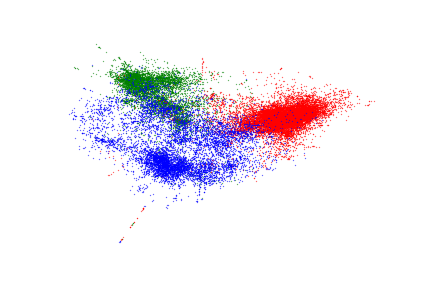}
\caption{DeepWalk}
\label{fig:subim1}
\end{subfigure}
\begin{subfigure}{0.18\textwidth}
\includegraphics[scale=0.25]{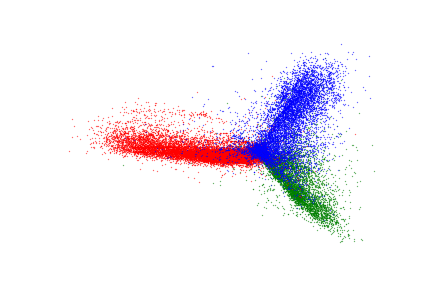}
\caption{Line}
\label{fig:subim1}
\end{subfigure}
\begin{subfigure}{0.18\textwidth}
\includegraphics[scale=0.25]{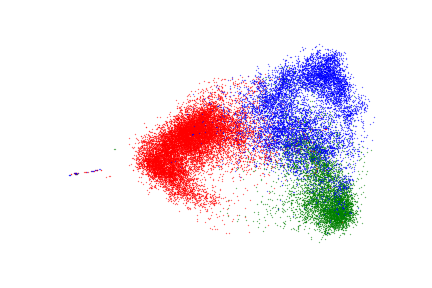}
\caption{node2vec}
\label{fig:subim1}
\end{subfigure}
\\
\begin{subfigure}{0.18\textwidth}
\includegraphics[scale=0.25]{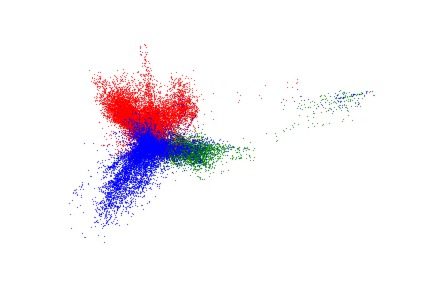}
\caption{TADW}
\label{fig:subim1}
\end{subfigure}
\begin{subfigure}{0.18\textwidth}
\includegraphics[scale=0.25]{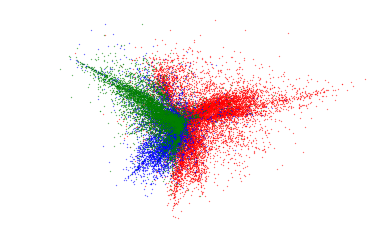}
\caption{AANE}
\label{fig:subim1}
\end{subfigure}
\begin{subfigure}{0.18\textwidth}
\includegraphics[scale=0.25]{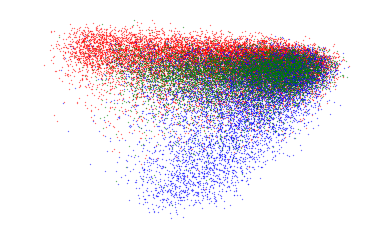}
\caption{Graphsage}
\label{fig:subim1}
\end{subfigure}
\begin{subfigure}{0.18\textwidth}
\includegraphics[scale=0.25]{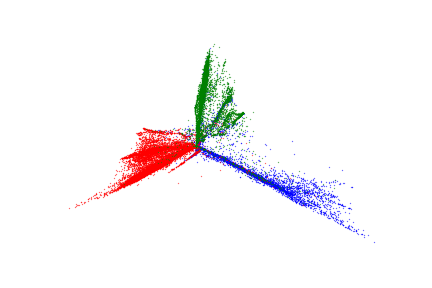}
\caption{FSCNMF++}
\end{subfigure}

\caption{Visualization of Microsoft Academia Graph dataset. Each point in the plot represents a node in the dataset. Each color corresponds to a particular group in the dataset.}
\label{fig:visuaMSA}
\end{figure*}

We have shown the visualization results for Pubmed-Diabetes and the MSA datasets in Figures \ref{fig:visuaPub} and \ref{fig:visuaMSA} respectively. Both the datasets have $3$ groups in each. We used all the data points from each of the datasets for the completeness of the plots. A close look into Figure \ref{fig:visuaPub} concludes that, though TADW and AANE use content along with structure to represent the network, they are not even able to show the existence of the three communities in this dataset. Whereas, FSCNMF++ is able to visually distinguish the communities well, and is slightly better that the closest baseline approach, which turns out to be node2vec in this case. For MSA dataset in Figure \ref{fig:visuaMSA}, FSCNMF++ is a clear winner for visualization, as the overlap between the three communities are minimum and they are well-separated compared to the visualizations by all the baseline approaches.

\begin{table}
\caption{F1 Score to measure the performance of different embedding algorithms for Multi-class classification on the Pubmed-Diabetes dataset. We used Random Forest classifier on the embeddings produced by each algorithm.}
\label{tab:multiclass}
\centering
\begin{tabular}{c|c|c c c c c}
\toprule
\multirow{2}{*}{Metric} & \multirow{2}{*}{Algorithm} & \multicolumn{5}{|c}{\centering  Train Size(\%)}\\
\cline{3-7}
& & {10} & {20} & {30} & {40} & {50}\\
\hline
\midrule
\multirow{5}{*}{Macro-F1} & Deepwalk & 69.60 & 72.78 & 74.64 & 75.82 & 76.79 \\
& LINE & 40.49 & 42.75 & 44.05 & 44.92 & 45.55 \\
& node2vec & 74.02 & 76.07 & 76.97 & 77.77 & 78.17 \\
& TADW & 73.45 & 75.91 & 77.03 & 77.79 & 78.40 \\
& AANE & 80.16 & 81.02 & 81.73 & 82.23 & 82.50 \\
& GrapSAGE & 65.36 & 68.77 & 70.63 & 72.05 & 73.29\\
& FSCNMF & \textbf{82.28} & \textbf{82.99} & \textbf{83.65} & \textbf{83.90} & \textbf{84.15} \\
\hline
\multirow{5}{*}{Micro-F1} & Deepwalk & 71.70 & 74.60 & 76.26 & 77.33 & 78.20 \\
& LINE & 44.40 & 46.42 & 47.40 & 48.15 & 48.74 \\
& node2vec & 75.71 & 77.48 & 78.29 & 79.03 & 79.36 \\
& TADW & 74.03 & 76.42 & 77.46 & 78.19 & 78.18 \\
& AANE & 79.98 & 80.85 & 81.59 & 82.08 & 82.41\\
& GraphSAGE & 68.55 & 71.43 & 72.93 & 74.11 & 75.11 \\
& FSCNMF & \textbf{82.46} & \textbf{83.24} & \textbf{83.19} & \textbf{84.13} & \textbf{84.43}\\
\bottomrule
\end{tabular}
\end{table}

\subsection{Applications to Multi-class Classification}
Multi-Class Classification involves using the embeddings generated by the algorithms to classify a given node into the category it belongs. Here the embedding are used as the features and the community which they belong to as the the true class labels. We have used Random Forest classifier to train the model to classify the embedding of a node into one of the classes.
We run the experiments on the Pubmed-Diabetes dataset for different training sizes ranging from 10\% to 50\%, repeating each such experiment 10 times.
Being consistent with the literature, we have also reported the average Macro-F1 and Micro-F1 scores to compare the performance of all the baseline algorithms over different training sizes in Table \ref{tab:multiclass}.

Results show the merit of FSCNMF, which is able to outperform all the baseline algorithms consistently over different training sizes.
We also observed that FSCNMF++ of order 1 (which is same as FSCNMF) performed the best among the higher orders of the same algorithm, hence reported the same. AANE, TADW and node2vec also perform well among the baseline algorithms. We observed similar results on the other three datasets also, which we are omitting because of the space limitation.

\begin{figure}[h]
\centering
\includegraphics[scale=0.25]{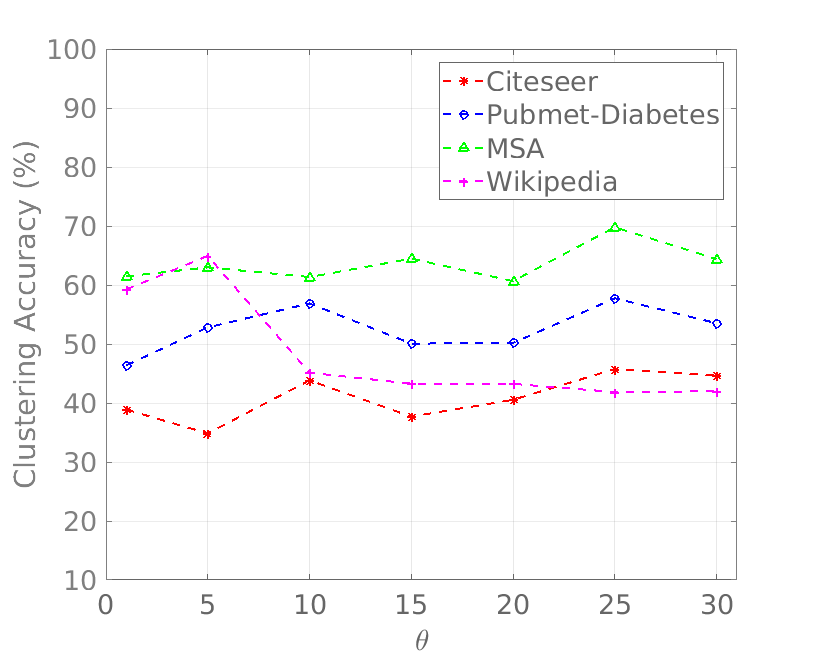}
\caption{Effect of the dimensionality ($\mathbf{k}$) of the embedding subspace on the performance of FSCNMF for node clustering. Here we assume $k = \theta \times \#communities$, where $\theta$ is a positive integer to control the value of $k$.}
\label{fig:dimen}
\end{figure}

\begin{figure}[h]
\centering

\begin{subfigure}{0.45\columnwidth}
\includegraphics[scale=0.22]{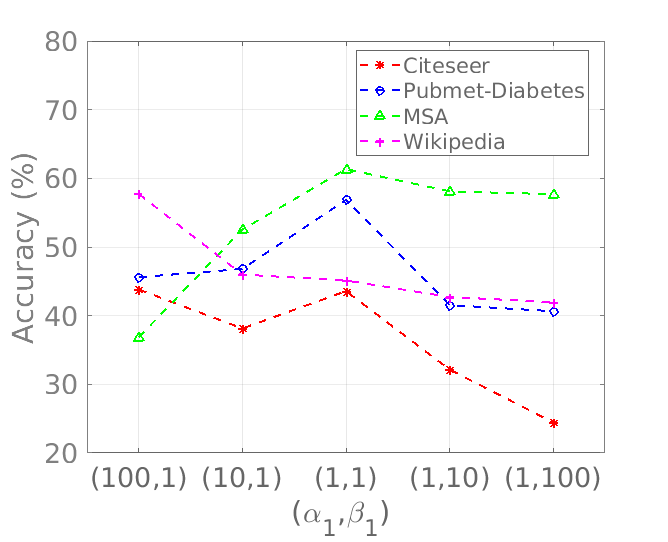}
\caption{}
\label{fig:subim1}
\label{fig:alpha_beta}
\end{subfigure}
\begin{subfigure}{0.45\columnwidth}
\includegraphics[scale=0.22]{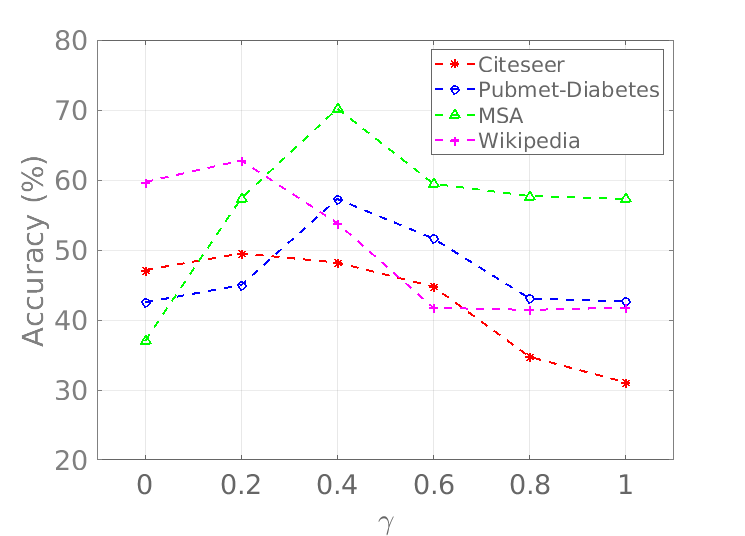}
\caption{}
\label{fig:subim1}
\label{fig:gamma}
\end{subfigure}

\caption{Clustering accuracy of FSCNMF by changing the preference over Content and Structure: (a) with different values of $(\alpha_1, \beta_1)$ in FSCNMF optimization (Section \ref{sec:FSCNMF}), (b) with different values of $\gamma$ for the Convex Combination of $B_1$ and $U$ to compute the final embedding  }
\end{figure}



\subsection{Parameter Sensitivity and Preference} \label{sec:sensitivity}
We conduct three different experiments in this section. First, we see the effect of the dimensionality (denoted by $k$, in Section \ref{sec:prob}) of the embedding subspace on the performance of FSCNMF. As all the datasets in our experiments have associated ground truth communities, we assume the dimension of the embedding subspace to be $k = \theta \times \#communities$, where $\theta$ is a positive integer. We vary $\theta$ and check the performance of the FSCNMF for node clustering on all the datasets, as shown in Figure \ref{fig:dimen}. It turns out that there is no single value of $\theta$ which can guarantee the best performance irrespective of any particular network. So further analysis is required to optimally find the best dimension of the embedding subspace for a given network, which can be addressed in future. With increasing $\theta$ time to inverse the $k \times k$ matrix would also increase (Section \ref{sec:scalability}).

Next, to check the effect of biasing the approach over structure or content, we experiment with different values of $\alpha_1$ and $\beta_1$ (in Eq. \ref{eq:cost1} and \ref{eq:cost2} respectively). With increasing $\alpha_1$, content is given more importance, as it drives the matrix $B_1$ to be similar to $U$, so that the component $\alpha_1 ||B_1 - U||_F^2 $ gets minimized in Eq. \ref{eq:cost1}. Similarly with increasing $\beta_1$, structure gets more importance. The performance of FSCNMF for different pairs of ($\alpha_1$, $\beta_1$) is shown in Fig. \ref{fig:alpha_beta}.

In the last paragraph, the biasness comes before the optimization of FSCNMF starts. But from the Section \ref{sec:FSCNMF}, another way to set the preference is to vary the value of $\gamma$, $0 \leq \gamma \leq 1$, so that the final embedding $\left( \gamma \times B_1 + (1-\gamma) \times U \right)$ can be biased to one of structure and content. Clearly when the value of $\gamma$ is close to 0, we give more importance to regularized content embedding matrix $U$ than the regularized structure embedding matrix $B_1$. The reverse happens when $\gamma$ is close to 1. To see this effect, we plotted the clustering accuracy of FSCNMF for different values of $\gamma$ in Figure \ref{fig:gamma}.

Interestingly, Figures \ref{fig:alpha_beta} and \ref{fig:gamma} are consistent. From these plots it is clear that for Citeseer and Wikipedia datasets, content is more informative than structure. Whereas for MSA, structure is better (same can be observed even in Section \ref{sec:clustering}), and for Pubmed-Diabetes the best result can be achieved when equal importance is given to both structure and content.

\section{Conclusion and Discussion}\label{sec:con}
In this paper, we have proposed a novel framework to fuse the link structure and the content of a network to generate the embeddings. 
We also develop fast update rules to optimize the associated cost functions. 
Experimental results show that our framework can successfully leverage the coherence of structure and content for multiple machine learning tasks on the network. Playing with different parameters of the framework, we are able to get key insight about the role of structure and content for different datasets.

The smaller runtime taken by spectral clustering algorithm on the embedding generated by FSCNMF needs to be explored further. One reason for this could be that the embedding generated by the FSCNMF is able to capture the underlying community structure of the network. Further, our framework is based on matrix factorization. So, naturally there is a critical trade-off between the use of disk I/O and main memory, which affects the runtime of the algorithms. Hence one can try to design an online version of the same, using approaches similar to those proposed in \cite{saha2012learning}.

Exploiting structure and content to generate an appropriate embedding is quite challenging. If there is a total agreement between the structure and content then TADW or its variants can do a good job. However, in real-life situations, there could be some noise or inconsistency between content and structure. In such situations, FSCNMF is the right choice as it offers a flexible framework to accommodate noise or partial inconsistency between structure and content (refer to Table\ref{tab:cluster}). This flexibility helps it to perform well even when the structure and content are totally consistent.

\bibliographystyle{ACM-Reference-Format}
\bibliography{FSCNMF}

\end{document}